\documentstyle[a4wide,12pt,epsf]{article}
\begin{document}
\begin{flushright}
\baselineskip=12pt
{RHCPP00-1T}\\
{astro-ph/0002391}\\
{MARCH 2000}
\end{flushright}

\begin{center}
{\Large \bf A New Dark Matter Model for Galaxies
\\}
\vglue 0.35 cm
{ 
George V. Kraniotis $^{\spadesuit}$ \footnote
 {G.Kraniotis@rhbnc.ac.uk}and Steven B. Whitehouse$^{\spadesuit}$ \footnote{
Sbwphysics@aol.com}\\}

{$\spadesuit$ {\it 
Centre for Particle Physics,\\
Royal Holloway College, University of London, \\
Egham, Surrey, TW20-0EX, U.K\\}}
\baselineskip=12pt

\vglue 0.25cm
ABSTRACT
\end{center}
{\rightskip=3pc
\leftskip=3pc
\noindent
\baselineskip=20pt
In this paper a new theory of Dark Matter is proposed.
Experimental analysis of several Galaxies show how the 
non-gravitational contribution to galactic Velocity Rotation 
Curves can be  interpreted as that due to the Cosmological Constant $\Lambda$. 
The experimentally determined values 
for $\Lambda$ are found 
to be consistent with those expected from Cosmological Constraints.
The Cosmological Constant is interpreted as leading to a constant energy 
density which in turn can be used to partly address 
the energy deficit problem (Dark Energy) of the Universe.
The work presented here leads to the conclusion that the 
Cosmological Constant is negative and that the universe is de-accelerating.
This is in clear contradiction to the Type Ia Supernovae results which support 
an accelerating universe.}

\vfill\eject
\setcounter{page}{1}
\pagestyle{plain}
\baselineskip=14pt

\section{Dark Matter}

The problem of missing or Dark Matter, namely that there is insufficient 
material in the form of stars to hold galaxies and clusters together, 
has been known since the pioneering work of Bessel, Zwicky and most recently 
Rubin \cite{VERA}.\\
\\
The existence of non-luminous 
Dark Matter was first inferred in 1984 by Fredrich Bessel from 
gravitational effects on positional measurements of Sirius and 
Procyon.  In 1933, Zwicky concluded that the velocity dispersion in Rich 
Clusters of galaxies required 10 to 100 times more mass to keep them 
bound than could be accounted for by luminous galaxies themselves.\\
\\
Finally, Trimble \cite{TRIMBLE} noted that the majority of galactic rotation curves, 
at large radii, remain flat or even rise well 
outside the radius of the luminous astronomical object.\\
\\
The missing Dark Matter has been traditionally explained in terms of
Dark Matter Halo's \cite{SCIAMA}, although none of the Dark Matter
Halo models have been very successful in explaining the experimental 
data \cite{ITALY,VAN}.\\
\\
This paper will describe the missing matter (Dark Energy) in terms of a
Cosmological Constant which leads to a constant energy density.\\
\\ 
The experimental determination of galactic velocity rotation curves (VRC) has 
been
one of the most important approaches used to estimate the
"local" mass (energy) density of the Universe. Several sets of data from
VRC's  will be analysed and the contribution due to the Cosmogical 
Constant determined.\\

\section{Constaints on the value of the Cosmological Constant}
It is interesting to estimate the allowed range of values for the 
Cosmological Constant within the constraints of General Relativity 
and observational astronomy, (for a comprehensive review, see 
Bahcall et.al. \cite{PAUL}).\\
\\
Starting from a General Relativity point of view, the 
Friedman energy equation is given by:
\begin{equation}
1=\frac{8 \pi G_N}{3}\frac{\rho_{matter}}{H^2}-\frac{k c^2}{R^2 H^2}+
\frac{c^2 \Lambda}{3 H^2},
\label{FRIEDMAN}
\end{equation}\\
where the Hubble Constant is denoted by $H$, 
the curvature term by  $k$ and
$G_N$ denotes the Newton gravitational constant.
Eq.(\ref{FRIEDMAN})  can be rewritten as
\begin{equation}
1=\Omega_{m}+\Omega_k+\Omega_{\Lambda}
\end{equation}
Here the relative contributions to the energy density of the universe 
are given by, the mass, curvature and Cosmological Constant.\\
\\
If we assume that the curvature contribution is 
small:
\begin{equation}
1=\Omega_{Matter}+\Omega_{\Lambda}
\label{TTT}
\end{equation}\\
\\
In order to satisfy equation (\ref{TTT}), it was surprising to 
discover that only a narrow range of values for the 
observed Cosmological Parameters were allowed.
A "reasonable" set of parameters consistent with observation are:
\begin{equation}
H_O=100Km s^{-1} Mpc^{-1}, \;
\rho_{matter}=5 \times 10^{-30} g cm^{-3}, \;
\frac{\Omega_{\Lambda}}{\Omega_{matter}}=4.3 
\end{equation}\\
\\
and $\Omega_{Matter}+\Omega_{\Lambda}=1.4$ (here we assume a 
small value for the curvature $\sim 0.4$. (For an
authoritative  review of the matter/energy sources of the universe, 
see Turner \cite{turner}).\\
\\
It was found that observational constraints placed upon the range of 
values 
for the  cosmological parameters lead to a surprisingly narrow 
range of possible values for the Cosmological Constant, the range 
being given by:
\begin{equation}
10^{-56} < |\Lambda| < 5\times10^{-55} cm^{-2}.
\end{equation}\\
\\

\section{Experimental Results}


\begin{table}
\begin{center}
\begin{tabular}{|c|c|c|}\hline\hline
${\rm Galaxy}$ & ${\rm Radius}$ 
& ${\rm Cosmological\; Constant}$  
\\ \hline\hline
NGC 2403&20Kpc  & $\Lambda_{NGC 2403}=3.6\times 10^{-55}cm^{-2}$  \\ \hline
NGC 4258&50 Kpc & $\Lambda_{NGC 4258}=5.5\times 10^{-55}cm^{-2}$ \\ \hline
NGC 5033&40 Kpc  & $\Lambda_{NGC 5033}=1.0 \times 10^{-55}cm^{-2}$  \\ \hline
NGC 5055&50 Kpc  & $\Lambda_{NGC 5055}=1.4\times 10^{-55}cm^{-2}$  \\ \hline
NGC 2903& 30 Kpc &$\Lambda_{NGC 2903}=3.8\times 10^{-55}cm^{-2}$ \\ \hline 
NGC 3198& 50 Kpc &$\Lambda_{NGC 3198}=5.0\times 10^{-56} cm^{-2}$ \\ \hline
\hline\hline
\end{tabular}
\end{center}
\caption{Absolute values of  the Cosmological Constant 
are shown above.}
\end{table}

It was shown \cite{SWGK}, within the Weak Field Approximation, 
that the Cosmological Constant at large radii could be determined 
from galactic velocity 
rotation curves. This contribution is given by:
\begin{equation}
v^2_{\Lambda}(r)=v^2_{obs}(r)-v^2_{mass}(r), {\rm \;\;\;leading\; to,} 
\label{vel}
\end{equation}
\begin{equation}
v^2_{\Lambda}/r=\frac{c^2 \Lambda r}{3},\;\;\;
{\rm at \;large \;r}
\end{equation}
The results obtained by this analysis are shown in Table 1.\\
\\
The experimental values obtained for the Cosmological Constant 
fall within the range determined from General Relativity 
and observational constraints.  While the initial results are promising, 
a thorough and systematic analysis of galactic rotation curves 
needs to be undertaken in order to confirm the trend.\\
\\
Previous results \cite{SWGK} reported for the value of the 
Cosmological Constant were 100 to 1000 times the "allowed value".  This 
systematic error arose for two main reasons:  the first by not taking 
the gradient of the curves at sufficiently large radii and the second by 
the lack of access to "real" experimental data leading to 
crude data analysis.\\
\\
The results presented in this paper suffer from the second problem, i.e. all 
the gradients were obtained from the data in the 
published literature and not from raw experimental data
i.e. M33 Corbelli $\&$ Salucci \cite{ITALY}, NGC 3198 \cite{HOL} and all 
others from \cite{KENT}. \\
\\
However 
experience has taught us that a cursory look at 
rotation curves will determine which galaxies are candidates for 
explanation by a Cosmological Constant and which are not. 
Galaxies where the velocity rotation curve remains flat or rises 
at large radii, are immediate candidates.
NGC 3198 is a good example, 
whereas others such as M33 \cite{ITALY} has clearly not relaxed to 
the Cosmological background, even at many times the galactic radii. A 
full explanation for M33 has to be sought in a different direction.\\
\\
Finally, a simple calculation of the effective mass density due to the 
Cosmological Constant in NGC 3198, 
\begin{equation}
\rho_{eff}=-\frac{c^2 \Lambda}{4\pi G_N}
\end{equation}
leads to a value of $5.4\times 10^{-29} g cm^{-3}$ which is 
comparable to the HI mass density  \cite{PAOLO} 
at the outer disk of NGC 3198 galaxy. This is further confirmation 
that the Cosmological Constant effect can be seen at galactic scale 
lengths.

\section{Accelerating or Decelerating Universe?}

Recently there has been great interest in the Type Ia Supernovae results 
of Perlmutter et al \cite{SUPERN} which suggest that the universe 
is accelerating.\\
\\
In this section we will show that the Weak Field Approximation 
coupled with galactic velocity rotation curve data inevitably lead to a 
negative Cosmological Constant.\\
\\
The equation for the VRC is given  \cite{SWGK} by
\footnote{In Ref.\cite{SWGK} eq.(15), there was a typographical sign error 
for one of the terms and also the negative pressure effect associated with 
$\Lambda$ was not fully appreciated.}
\begin{equation}
-\frac{v^2}{r}=-\frac{G m}{r^2}+\frac{c^2 \Lambda}{3} r
\label{improv}
\end{equation}
We note that Eq.(\ref{improv}) is only strictly true for small and 
large radii, however it will serve to illustrate our 
arguments.\\
\\
Using the Newtonian limit 
of Einstein field equations we derived 
equation (\ref{improv}). It is important to realize that 
the Cosmological Constant obeys the equation of state given by,  
\begin{equation}
P_{\Lambda}=-c^2 \rho_{\Lambda},
\end{equation}
where $P_{\Lambda}$ is the pressure term due to $\Lambda$.
Taking the Newtonial limit in the 
absence of matter, $T_{\mu\nu}=0$, the differential equation for 
the static 
Newtonian potential becomes
\begin{equation} 
\nabla^2 \Phi=-c^2 \Lambda
\label{potential}
\end{equation}
leading to, 
\begin{equation}
\rho_{eff}=\rho_{\Lambda}+\frac{3 P_{\Lambda}}{c^2}=-2\rho_{\Lambda}
\end{equation}
\\
If we arbitrary set $\Phi=0$ at the origin, then in spherical 
coordinates (\ref{potential}) has the solution 
$\Phi=-\frac{c^2 \Lambda}{6}r^2$ \cite{SIROHA}. Thus, the Cosmological 
Constant leads 
to the following correction to the Newtonial potential 
\begin{equation}
\Phi=-\frac{Gm}{r}-\frac{c^2 \Lambda}{6}r^2
\end{equation}
At small galactic radii the velocity versus radius contribution is 
well known and follows Newtonian physics. For large radii a negative 
Cosmological Constant gives  a positive contribution to the 
VRC which is what is actually observed.
On the other hand the effect of a positive Cosmological Constant  
would be  to lower the rotation curve below that due 
to matter alone. \\
\\
The above simple argument, based on observational astronomy, 
allows only a negative Cosmological Constant as a possible explanation  
for the galactic velocity rotation curve data.
This is in clear disagreement with the Type Ia supernovae results 
\cite{SUPERN}. However, given  the uncertainties in the determination 
of the deceleration parameter, $q_0$, derived from supernovae data 
\cite{SUPERN} the approach 
outlined  above has certain merits worth consideration. \\
\\
In summary these are ,  the 
Cosmological Constant is determined from $direct$ measurement 
unlike the Supernovae results, the experimentally determined value 
is  the correct 
order of magnitude as that required from cosmological constraints, and 
finally a negative Cosmological Constant is consistent, and indeed 
a natural physical explanation , for the observed galactic 
velocity rotation 
curve data.\\
\\
Finally, observations of global clusters of stars constrain 
the age of the universe and consequently place an observational limit 
on a negative Cosmological Constant \cite{OHA} of , 
\begin{equation}
|\Lambda| \leq 2.2 \times 10^{-56} cm^{-2}.
\label{limit}
\end{equation}
Note, the Cosmological Constant derived from 
global cluster constraints is in agreement with the 
experimentally determined value derived from 
galactic velocity rotation curve data.

\subsection{Experimental Tests-Dark Matter Halo vs Cosmological Constant }
It would be of some interest if it was possible to experimentally 
distinguish between the contribution of Dark Matter Halo's and Dark Energy 
(Cosmological Constant) to galactic rotation curves.\\
\\
We know that Dark Matter predicts a variation of mass at large radii given by
\cite{KOLBE},
\begin{equation}
M_{DM}(r)\propto r
\end{equation}
while for Dark Energy due to a Cosmological 
Constant,
\begin{equation} 
M_{\Lambda}(r)\propto r^3[\rho_{\Lambda}+(3 P_{\Lambda}/c^2)].
\end{equation}\\
\\
With these different types of predictive variations it should be 
possible to design experimental tests to distinguish between the 
two phenomena.

\section{Quark Hadron Phase transition}

In this section which is of more speculative nature, working 
within the Extended Large Number Hypothesis, and using the 
experimentally  determined Cosmological Constant, we will demonstrate  
how the energy density for the Quark - Hadron can be estimated.\\
\\
However, it is useful to put into context the significance of the 
Cosmological Constant for many seemingly disparate 
branches of Physics.  The figure 1 below shows the Cosmological Constant 
at the epicentre of Physics.\\
\\
The diagram demonstrates a dichotomy whereby several branches of Physics 
need a non-zero Cosmological Constant in order to explain key 
physical phenomena, whilst in others a non-zero value 
presents a fundamental problem.\\
\\
\begin{figure}[h]
\epsfxsize=4.2in
\epsfysize=3.3in
\epsffile{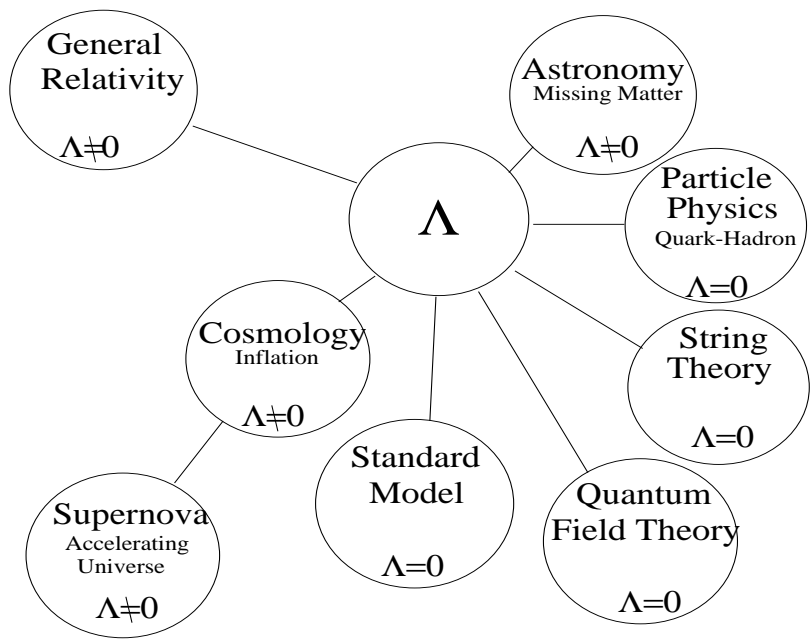}
\caption{$\Lambda$ at the epicentre of Physics}
\end{figure}

It is also noted here that while fundamental 
theories of Particle Physics such as 
 the Standard Model, Quantum Field Theory and 
String Theory have many major predictive successes they all have 
problem with a high vacuum energy density. 
On the other hand while the Extended LNH is formulated from a naive 
theory \cite{ZELDO} it appears to correctly predict 
the correct vacuum energy density and other cosmological parameters.
The Extended LNH relates the value of the Cosmological Constant to 
the effective mass given by:
\begin{equation}
|\Lambda|=\frac{G_N^2 m^6_{eff}}{h^4}=\frac{c^6 L_s^4}{h^6}m^{6}_{eff}
\label{constant}
\end{equation}
Matthews \cite{MATTHEWS} pointed out that when using the Extended LNH to 
determine today's cosmological parameters, the mass of the 
proton originally suggested by Dirac should be replaced by 
the energy density of the last phase transition of the 
Universe : Quark - Hadron.\\
\\
Note that in equation (\ref{constant}) there are no 
free parameters, $L_s$ is normalised to the 
gravitational constant and corresponds to 
the fundamental length of 
String Theory.\\
\\
Using equation (\ref{constant}) and the Cosmological Constant determined from NGC 5033, the effective mass is given by 
\begin{equation}
m_{Effective}=332 MeV
\end{equation}
We will associate this value with the Quark - Hadron phase 
transition energy.  (Other experimentally determined 
Cosmological Constant data give $m_{QH}$ in the range 295 - 410 MeV).
The experimentally determined value within the LNH predicts the 
correct order of magnitude for the phase transition.\\
\\
The above result  poses the question that it might be possible to gain 
insights on the quantum mechanical origin of the Universe, 
as Dirac \cite{NUMBER,DIRAC,LARGE} suggested, from direct observation of 
the present day Universe.\\
\\
Finally,  does the Cosmological Constant 
provide the key to the integration of the various Physics 
disciplines as Figure 1 suggests?\\
\\

\section{Discussion}

Analysis of the galactic rotation curves 
show that the missing Galactic Dark Matter 
can be explained in terms of a Cosmological Constant.\\
\\
This contribution can be considered a prime 
candidate for the "Dark Energy" which is smoothly distributed 
throughout space, and 
contributes approximately $70\%$ to the mass/energy of 
the Universe \cite{turner}.\\
\\ 
However, in order to support this thesis for the Cosmological Constant, 
thorough and systematic analysis of galactic velocity rotation 
data needs to take place.\\
\\
It was shown how, within the Weak Field Approximation, 
that  VRC data inevitably 
lead to a negative value for the Cosmological Constant in direct 
disagreement with the type Ia Supernovae data. Nevertheless, given 
the uncertainties in determining the 
deceleration parameter $q_0$ \cite{HOYLE}, from the 
redshift-magnitude 
Hubble diagram using Type Ia supernovae as standard candles, we believe 
our approach is worth further consideration.\\
\\
The experimental values determined for the 
Cosmological Constant are shown to lie within an 
acceptable range. These values, used within the Extended Large 
Number Hypothesis, predict  values for 
the Quark-Hadron phase transition energy 
in the range 295-410 MeV.\\
\\
It would be remarkable, if proved correct, that the 
Cosmological Constant could be directly 
determined from the analysis of galactic velocity rotation curves.\\
\\
Equally remarkable, if proved correct, is the idea 
that astronomical observations can shed light on the 
last quantum mechanical phase transition of the Universe, namely the 
Quark-Hadron.\\
\\

\section{Acknowledgements}

We would like to thank  Paolo Salucci for 
invaluable discussions on 
Cosmological and Astronomical aspects related to this work and Alexander 
Love for useful comments on the manuscript. We also thank J. Hargreaves 
and D. Bailin for suggestions and 
useful discussions.\\
\\
We also would like to thank Deja Whitehouse for proof 
reading this document.\\
\\
George Kraniotis was supported for this work by PPARC.


\end{document}